\documentclass[aps,prd,floatfix,preprint,nofootinbib]{revtex4}

\usepackage{amsmath,amssymb,graphicx,color}
\usepackage{hhline}

\newcommand{\dip}{Dipartimento di Fisica, Universit\`a di Trento\\
                                     via Sommarive 14, 38123 Trento, Italia\\}
\newcommand{\infn}{TIFPA (INFN)\\via Sommarive 14, 38123 Trento, Italia\ \medskip}
\newcommand{\maxR}[1]{Massimiliano Rinaldi${}^{#1}$\footnote{e-mail:\sl massimiliano.rinaldi@unitn.it\rm}}
\newcommand{\guido}[1]{Guido Cognola${}^{#1}$\footnote{e-mail:\sl cognola@science.unitn.it\rm}}

\newcommand{\luciano}[1]{Luciano Vanzo${}^{#1}$\footnote{e-mail:\sl vanzo@science.unitn.it\rm}}
\newcommand{\sergio}[1]{Sergio Zerbini${}^{#1}$\footnote{e-mail:\sl zerbini@science.unitn.it\rm}}

\DeclareMathOperator\im{Im}
\newcommand{\be}{\begin{equation}}
\newcommand{\ee}{\end{equation}}
\newcommand{\beq}{\begin{eqnarray}}
\newcommand{\eeq}{\end{eqnarray}}
\newcommand{\non}{\nonumber}
\begin{document}

\title{Thermodynamics of topological black holes in $R^{2}$ gravity}

\author{\guido{}} \author{\maxR{}}  \author{\luciano{}} \author{\sergio{}}
%\email{massimiliano.rinaldi@unitn.it}
\affiliation{\dip}
\affiliation{ \infn}

\date{\today}
%%%%%%%%%%%%%%%%%%%  ABSTRACT %%%%%%%%%%%%%%%

\begin{abstract} 
\noindent We study topological black hole solutions of the simplest quadratic gravity action and we find that two classes are allowed. The first is asymptotically flat and mimics the Reissner-Nordstr\"om solution, while the second is asymptotically de Sitter or anti-de Sitter. In both classes, the geometry of the horizon can be spherical, toroidal or hyperbolic. We focus in particular on the thermodynamical properties of the asymptotically anti-de Sitter solutions and we compute the entropy and the internal energy with Euclidean methods. We find that the entropy is positive-definite for all horizon geometries and this allows to formulate a consistent generalized first law of black hole thermodynamics, which keeps in account the presence of two arbitrary parameters in the solution. The two-dimensional thermodynamical state space is fully characterized by the underlying scale invariance of the action and it has the structure of a projective space. We find a kind of duality between black holes and other objects with the same entropy in the state space. We briefly discuss the extension of our results to more general quadratic actions.
\end{abstract}
 
 %%%%%%%%%%%%%%%%%%%%%%%%%%%%%%%%%%%%%%%%

\maketitle

\section{Introduction}

\noindent Quadratic gravity has always attracted a lot of interest for its classical scale-invariant properties. Recently,  scale invariant quadratic models were reconsidered within a much  larger physical context in \cite{strumia}, where it was shown that they lead to an inflationary  model consistent with observations, provided one adds a new scalar field degree of freedom and takes in account the running of  the coupling parameters. In this context, a much simpler inflationary scenario, consistent with the latest observational data, was presented in \cite{tn}.

Quadratic gravity models are particularly attractive as it is believed  that they are renormalizable and asymptotically free \cite{stelle}-\cite{buch}, although ghosts are in general present. However,  the simplest case of $f(R)=R^{2}$ is also ghost-free \cite{kou}. De Sitter and anti-de Sitter black hole solutions of this kind of theory were investigated and their thermodynamical properties discussed in pioneering works, such as \cite{Jaco,miriam}. Furthermore, 
inflation in quadratic gravity models, accounting for running coupling constants, was recently studied also in \cite{sergei14}. A related black hole instability was investigated in modified gravity in \cite{Seba}.

The peculiar properties of these theories stem from the underlying scale invariance of the action, which forbids the presence of any length scale, in contrast to general relativity (GR), where the vacuum action contains the Planck mass and, eventually, a dimensionful cosmological constant. Thus, for example, one cannot tell the frequency of a gravitational wave in this theory, but the ratios of the frequencies of two waves is meaningful. Similarly, by observing which events can be reached from a given event by a gravity wave, or some other massless signal, one can determine the local light cone. However, this is as far as one can go in general. Thus,  the space-time metric has no direct physical meaning, only ratios of intervals really matter\footnote{In the present discussion the role of the observer is much trickier  than usual. We assume that  a test body could be introduced without disturbing appreciably  the field, thus the concept of observations makes sense.}.  In the same way, masses or other scales are meaningless as soon as we manage to maintain the scale invariance symmetry, so we argue that to measure anything meaningful one has to break this symmetry. In principle, one can imagine detectors with no built-in scales.  For example, one can use as a clock a given gravitational wave of $R^2$ gravity, and use it to measure the ratio of its arbitrary frequency to that of other waves by observing interference phenomena, which are, in principle, calculable using the theory.   

We emphasize also the peculiar fact that scale invariance is extremely sensitive to external perturbations. Even the coupling of a very low mass particle or the introduction of a clock network with its necessary built-in scale, will break the scale symmetry substantially \footnote{This aspect was already mentioned a long time ago in \cite{sexl:1966}.}. Indeed, according to a very general theorem, for asymptotically flat initial data the total energy in $R^2$-gravity is exactly zero \cite{Boulware:1983td}, but any other material device coupled to the field will have positive energy. For de Sitter or anti-de Sitter boundary conditions the situation is different. We will see that black hole thermodynamics suggests that a continuous mass spectrum is possible. This is in accord with the general understanding of the scale invariance symmetry, according to which it requires only dimensionless coupling constants and either no masses or a continuous mass spectrum\footnote{Since $-m^2=P^2$, the dilation generator $D$ changes the mass  via $\delta P^2=\lambda[D,P^2]=-\lambda P^2$.}.

In this paper we wish to explore spherically symmetric solutions, with topological horizon, along the lines of the early work of Buchdahl \cite{buchdahl}. We  will consider the simplest quadratic model $f(R)=R^{2}$, and we will generalize the solutions with spherical horizon found in \cite{deser03} and reconsidered recently in \cite{riotto}. We will pay special attention to the thermodynamical properties of the asymptotically anti-de Sitter solutions, as they clearly display the underlying scale invariance of the action.

In the next section we derive the general vacuum solutions of $R^{2}$ gravity with spherical symmetry and topological horizon. In Sec.\  3 we study the thermodynamical properties of these solutions and interpret the results from the point of view of the underlying scale invariance symmetry.  We conclude in Sec.\  4 with a discussion of our results and their extensions to more general quadratic actions.

\section{Topological black holes in $R^{2}$ gravity}

\noindent Let us first  consider the generic modified gravity Lagrangian
\beq
{\cal L}=\sqrt{g}f(R)\,,
\eeq
for which the equations of motion read \cite{defelice, sergeirev}
\beq
XR_{\mu\nu}-{1\over 2}fg_{\mu\nu}-\nabla_{\mu}\nabla_{\nu}X+g_{\mu\nu}\square X=0\,,
\eeq
where we set
\beq
X={df(R)\over dR}\,.
\eeq
Let us now now restrict to the case $f(R)= R^{2}$. The equations simplify to
\beq\label{eom}
2RR_{\mu\nu}-\frac12 R^{2}g_{\mu\nu}-2\nabla_{\mu}\nabla_{\nu}R+2g_{\mu\nu}\square R=0\,, 
\eeq
while the trace of this equations reduces to $\square R=0$. In this paper we choose to work in Jordan frame only since the Einstein frame is potentially ill-defined. In fact, the conformal transformation to the Einstein frame reads $g_{\mu\nu}\rightarrow \Omega^{2} g_{\mu\nu}$ with $\Omega^{2}=X$. Hence, all the solutions with $R=X=0$  are excluded from the conformal mapping. Given that a class of spherically symmetric solution with identically vanishing Ricci scalar exists (see below), we prefer to work in the Jordan frame.

We now look for spherically symmetric solutions with metric
\beq
ds^{2}=-e^{2N(r)}dt^{2}+e^{-2N(r)}dr^{2}+r^{2}d\Sigma^{2}_{k}\,,
\label{sst}
\eeq
where
\beq
d\Sigma^{2}_{k}={d\rho^{2}\over 1-k\rho^{2}}+\rho^{2}d\phi^{2}\,,\quad k=0,\pm1\,,
\eeq
parametrizes the geometry of the horizon ($k=1$ spherical, $k=0$ flat or toroidal, $k=-1$ hyperbolic).
With this metric ansatz we have three differential equations of third or fourth order for $N(r)$. By defining
\beq
N_{1}\equiv{dN\over dr},\quad N_{2}\equiv{d^{2}N\over dr^{2}},\quad N_{3}\equiv{d^{3}N\over dr^{3}},\quad N_{4}\equiv{d^{4}N\over dr^{4}}\,,
\eeq
the system of equations reads
\beq
&&\Big[   7-(2N_1N_3+4N_{1}^4+8N_{2}N_{1}^2-N_{2}^2)r^4-(4N_{3}+24N_{1}^3+28N_{1}N_{2})r^3\\\non
&&-(16N_{2}+28N_{1}^2)r^2+8N_{1}r  \Big]e^{4N}-6ke^{2N}-k^{2}=0\,,\\\non
&&\Big[ 5+(20N_{1}^4+18N_{1}N_{3}+56N_{2}N_{1}^2+2N_{4}+11N_{2}^2)r^4+4(3N_{3}+19N_{1}N_{2}+14N_{1}^3)r^3\\\non
&&+4(N_{2}+N_{1}^2)r^2-8N_{1}r\Big]e^{4N}-6ke^{2N}+k^{2}=0\,,\\\non
&& \Big[  (76N_{2}N_{1}^2+36N_{1}^4+20N_{1}N_{3}+2N_{4}+13N_{2}^2)r^4+10(8N_{1}N_{2}+8N_{1}^3+N_{3})r^3\\\non
&& -4(N_{2}+4N_{1}^2)r^2-16N_{1}r+7\Big]e^{4N}+6k(2N_{1}r-1)e^{2N}-k^{2}=0\,.
\eeq
If we solve algebraically the system for $N_{2},N_{3}$ and $N_{4}$ we find two inequivalent solutions.

\subsection{Asymptotically (A)dS black holes}

\noindent  The first class of solutions corresponds to the system
\beq
N_{2}&=&-{k-e^{2N}+2N_{1}^{2}r^{2}e^{2N}\over r^{2}e^{2N}}\,,\\\non
N_{3}&=&{2\left(4e^{2N}N_{1}^{3}r^{3}-2e^{2N}N_{1}r+3N_{1}kr-e^{2N}+k\right)\over e^{2N}r^{3}}\,,\\\non
N_{4}&=&-{2\over r^{4}}\left[   24N_1^{4}r^4-16N_{1}^{2}r^{2}-4N_1r -1+2k(2N_{1}r+1)(6N_{1}r-1)e^{-2N} +3k^{2}e^{-4N}  \right]\,.
\eeq
The first of these equations is a second order differential equation that can be solved, yielding
\beq\label{classA}
N(r)={1\over 2}\ln \left(k+{a\over r}+br^{2}\right)\,,
\eeq
with $a$ and $b$ arbitrary constant. The other two equations are identically satisfied by this solution, so the system is consistent. The structure of the spacetime depends upon combinations of $k$, $a$, and $b$. In particular,  the zeros of the function $-g_{tt}=\exp(2N)$ determine the location of the horizons. Here, we adopt the mostly plus signature, therefore we require $a<0$ so that, when $r\rightarrow 0^{+}$, $g_{tt}\rightarrow \infty$ independently of $k$ and $b$. In accordance with usual GR notation, we set $\Lambda=-3b$  and we  write \footnote{As a reference, we assume that the GR Lagrangian has the form $L=(m_{p}^2/2)(R-2\Lambda)\sqrt{g}$,  $m_{p}$ being the Planck mass.}
\beq\label{gtt2}
-g_{tt}=k-{\omega M\over r}-{\Lambda r^{2}\over 3}\,,
\eeq
In GR, $\omega$ is a parameter that depends on the volume of the horizon space per unit radius and on the Newton constant.  However, in quadratic gravity,  we have no natural Planck scale, so we need to keep in mind that $\omega M$ really represents a length scale, if we agree with conventions that coordinate differentials  are to represent lengths.

According to the value of $k$, we have  three cases:
\begin{description}
\item[$k= 1$ :] for $\Lambda<0$ we have a single horizon and the well-known  Schwarzschild anti-de Sitter black hole; for $\Lambda>0$ we have the Schwarzschild - de Sitter black hole with two unstable horizons or the single horizon Nariai black hole; in both cases $\omega=4\pi$.
\item[$k=0$ :] there exists a unique horizon  at $r_{+}=-12M\Lambda^{-1}$ provided $\Lambda<0$; choosing a Teichm\"{u}ller parameter $\tau$ to specify the conformal class of the torus \cite{Vanzo:1997gw}, we may set $\omega=|\im\tau|$.
\item[$k=-1$ :] there exists a unique horizon for $\Lambda<0$, and $\omega=-2\pi\chi_{g}$, with $\chi_{g}=2-2g$ the Euler number of the horizon manifold.
\end{description}
For $k=-1,0$ the metrics are usually dubbed topological black holes \cite{Aminneborg:1996iz,Cai,mannTBH,Brill:1997mf}. The curvature invariants are independent of $k$ as the horizon is an Einstein space \cite{Vanzo:1997gw,topbh} and read
\beq
R=36\Lambda,\quad R_{\mu\nu}R^{\mu\nu}=324\Lambda^{2},\quad R_{\mu\nu\alpha\beta}R^{\mu\nu\alpha\beta}={12(\omega^{2}M^{2}+18\Lambda^{2}r^{6})\over r^{6}}.
\eeq
From the last expression, we see that a physical singularity appears at $r=0$ for any non-vanishing value of $\Lambda$ and $M$.

\subsection{Asymptotically flat black holes}

\noindent  The second, inequivalent, class of solutions is obtained from 
\beq
N_{2}&=&{k-e^{2N}-4e^{2N}N_{1}r-2N_{1}^{2}r^{2}e^{2N}\over r^{2}e^{2N}}\,,\\\non
N_{3}&=&{2\left(4e^{2N}N_{1}^{3}r^{3}+12e^{2N}N_{1}^{2}r^{2}-3N_{1}kr+12N_{1}re^{2N}+3e^{2N}-3k\right)\over e^{2N}r^{3}}\,,\\\non
N_{4}&=&-{6\over r^{4}}\left[  8N_{1}^{4}r^{4}+32N_{1}^{3}r^{3}+48N_{1}^{2}r^{2}+32N_{1}r+7 -8k(N_{1}r+1)^{2}e^{-2N}+k^{2}e^{-4N} \right]\,.
\eeq
Again, the first equation can be solved exactly and one finds that
\beq
N(r)={1\over 2}\ln\left(k+{a\over r}+{b\over r^{2}} \right)\,.
\eeq
As before, the other equations are identically satisfied by this solution, which looks like a Reissner-Nordstr\"om black hole, for which $R=0$ and
\beq
R_{\mu\nu}R^{\mu\nu}={4b^{2}\over r^{8}},\quad R_{\mu\nu\alpha\beta}R^{\mu\nu\alpha\beta}={4(3a^{2}r^{2}+12abr+14b^{2})\over r^{8}}\,.
\eeq
Again, the black hole horizon exists depending upon combinations of $k$, $a$ and $b$ according to the following scheme  \footnote{Of course, we should not claim asymptotic flatness for all $k$, but $k=1$.}:
\begin{description}
\item[$k=1$ :] the horizon  exists provided  $(a>0,b<0)$ or $(a<0,b\leq a^{2}/4)$; it is formally identical to the charged Reissner-Nordstr\"om family of solutions; its analytic extension contains infinitely many time-like naked singularities alternating with Killing horizons;
\item[$k=0$ :]  the  horizon is located at $r_{+}=-b/a$ so it is well-defined for $(a<0,b>0)$ or $(a>0,b<0)$; however, for $-m=a<0$ and $q^2=b>0$, the metric is dynamical at large positive $r$ while the static region encloses a naked singularity, so it looks like a reversed black hole and it is not asymptotically flat. There is also another asymptotically flat region bounded by the singularity for negative $r$, with no horizons there, which seems disconnected from the first unless one can glue manifolds along the singularity. The other case is a black hole enclosing a space-like singularity, so the causal structure appears to be a toroidal version of the Schwarzschild solution; 
\item[$k=-1$ :]  the horizon  exists provided  $(a<0,b>0)$ or $(a>0,b>-a^{2}/4)$. However in both cases the metric is dynamical at large $r$, so it can hardly be said to represent a black hole.  Instead there is a static region enclosed between the horizons, and continuation to negative values of $r$ seems possible. The largest zero looks like a cosmological horizons, inasmuch as it encloses the static region. A full determination of the causal structure will not be attempted here.
\end{description}
To summarize, for the asymptotically flat class of solutions, we shall  consider only spherical and toroidal black holes, the hyperbolic ones deserving special considerations that go beyond the present paper.

\section{Thermodynamics}

\subsection{Asymptotically AdS black holes}\label{ssA}

\noindent The thermodynamical properties of topological black holes in GR can be defined in terms of the Euclidean action $I_{E}$. If the latter is finite and positive-definite, one can construct the tree level partition function $Z=\exp(- I_{E})$ and formally define the internal energy and the entropy as it is done in the canonical ensemble \cite{Gibbons:1976ue}
\beq\label{thermo}
E={\partial I_{E}\over\partial\beta}, \quad S=\beta E-I_{E}\,,
\eeq
where $\beta$ is the periodicity of the Euclidean black hole metric, interpreted as the inverse of the horizon temperature. Technically, the Euclidean action should be implemented by boundary terms \cite{GHYterm}, which, however, do not contribute to the thermodynamical quantities for either spherical or topological black holes in GR \cite{hptrans,topbh,Vanzo:1997gw}. As we will briefly recall below, in $f(R)$ gravity boundary terms can be important \cite{boundary}. However, for the case at hand they turn out to be irrelevant, just like in GR. 

The thermodynamics of black holes in $f(R)$ theories crucially depends on the asymptotic form of the metric. On a very general ground \cite{wald}, it can be shown that if the black hole is asymptotically flat, then its entropy has the form \cite{boundary}
\beq\label{wald}
S=16\pi X(R_{H}) {A\over 4G}\,,
\eeq
where $A$ is the horizon area, $G$ is the Newton's constant, and $X(R_{H})$ denotes the derivative of $f(R)$ with respect to $R$, evaluated at the black hole horizon. 

For the class of asymptotically flat solutions this formula yields a vanishing entropy, since the $R=0$ everywhere. Therefore, these black holes seem to have non-vanishing temperature but zero entropy, as it will be discussed below.  The other class of black holes is asymptotically anti-de Sitter with an infinite Euclidean action. Therefore, one needs a subtraction procedure between the black hole solution and a suitable background. This method has been widely used in GR to consistently define the entropy of topological black holes in anti-de Sitter space \cite{Vanzo:1997gw,topbh}. We can apply the same method for the asymptotically anti-de Sitter black holes of $R^{2}$ gravity provided one interprets correctly the dependence of the internal energy from the parameters of the theory. We recall that in  GR the cosmological constant appears as a parameter in the action. Therefore, the only parameter that can be, in principle, varied is the black hole mass $M$ and, in fact, the internal energy is proportional to $M$ only \footnote{Technically, for $k=-1$ the internal energy depends on a critical mass value which is related to the cosmological constant. However, the entropy does not and the first law of thermodynamics $TdS=dE$ applies \cite{topbh}. }.

 In the case of $R^{2}$ instead, the radius of the anti-de Sitter space, defined in our conventions by $\ell^{2}=-3/\Lambda$, is arbitrary. Therefore, we expect that the entropy, the internal energy and so on, depend not only on $M$ but also on $\ell$.  This is not a mere expectation: as we will see, we are actually forced to vary $\ell$ as a direct consequence of scale invariance. Incidentally, the idea of considering the cosmological constant as a thermodynamical variable has a nearly 20-year old history. A comprehensive  recent analysis, with references, for anti-de Sitter black holes within Lovelock gravity  can be found in \cite{Frassino:2014pha}, for Lovelock-Born-Infeld gravity in \cite{Mo:2014qsa} and Gauss-Bonnet topological black holes in \cite{Cai:2013qga,Xu:2013zea}. As a further consequence of the new role for $\ell$, we expect that the usual first law is modified as well. We now show that these expectations are indeed correct. 
 
 Let us first write 
 \beq
 -g_{tt}=k-{ L\over r}+{r^{2}\over \ell^{2}}\,,
 \eeq
where $L=\omega M$ is now to be considered as an arbitrary length. Let us also denote with $r_{+}=r_{+}(L,\ell)$ the radius of the event horizon, defined as the largest zero of the equation $g_{tt}=0$. We associate an inverse temperature $\beta$ to the horizon according to the formula 
 \beq\label{bhbeta}
 \beta=-4\pi\left(dg_{tt}\over dr\right)^{-1}={4\pi\ell^{2}r_{+}\over 3r_{+}^{2}+k\ell^{2}}\,.
 \eeq
 It is known that, in GR and for $k=1$, the temperature has a minimum value, below which the black hole dissolves into pure radiation (a phenomenon referred to as the Hawking-Page transition, \cite{hptrans}). On the opposite, for $k=0,-1$ the black hole solution  dominates over the empty anti-de Sitter space at all temperatures so there is no phase transition.
For $k=-1$, if we require the temperature to be positive, we find that
\beq\label{rcrit}
r_{+}\ge r_{c}\equiv {\ell\over \sqrt{3}}\,,
\eeq
which corresponds to the critical length
\beq\label{mcrit}
L_{c}=-{2\ell\over 3\sqrt{3}}\,.
\eeq
This quantity is crucial in order to define the appropriate background to be subtracted from the Euclidean action when $k=-1$ (below we shall see that there is another possible background choice for any $k$).

Let us define the quantity 
\beq\label{subt}
\Delta I_{E}=I_{E}^{bh}-I_{E}^{bk}\,,
\eeq
where the first term denotes the Euclidean action for the black hole solution and the second for a suitable background space. In general, the Euclidean action for $F(R)$ gravity contains a bulk part and a boundary term of the form \cite{boundary}
\beq
I_{bound}\sim \oint d^{3}x\sqrt{h}F'(R)K\,,
\eeq
where $h$ is the determinant of the metric on the boundary and $K$ the trace of its extrinsic curvature. In our case, we have $X=2R=$ const (see eq.\ \eqref{eom}), therefore the difference between the background and the black hole terms reduces to the integral of the differences of the respective extrinsic curvatures at a large radius, where they coincide. Thus, the boundary term does not give any contribution to the expression \eqref{subt}.

We now compute the two contributions in eq.\  \eqref{subt}. The background term reads \footnote{The Euclidean continuation is performed in such a way that the Euclidean action is positive-definite, thus if the Lorentzian action is defined as $I_{L}=\int d^{4}xR^{2}$, the Euclidean one is obtained by multiplying it by $i$ together the continuation $t=i\tau$.}
\beq
I_{E}^{bk}=\int d^{4}x\sqrt{g}R^{2}=\int_{0}^{\beta_{0}}d\tau\int_{r_{c}}^{\bar r}drr^{2}R^{2}={48\beta_{0}\over \ell^{4}}(\bar r^{3}-r_{c}^{3})\,.
\eeq
The integration over $r$ spans the interval $[r_{c},\bar r]$, where $\bar r$ is some arbitrarily large radius. Here,  $r_{c}=0$ for $k=1,0$ while, for $k=-1$, it is defined by \eqref{rcrit}. Since this background corresponds to an empty space, the inverse temperature $\beta_{0}$ is arbitrary. However, for the subtraction \eqref{subt} to be consistent, $\beta_{0}$ must match with the inverse black hole temperature, defined by eq.\ \eqref{bhbeta}, at  $\bar r$. This is guaranteed if
\beq
\beta_{0}^{2}\left(k-{L_{c}\over \bar r}+{\bar r^{2}\over \ell^{2}}\right)=\beta^{2}\left(k-{L\over \bar r}+{\bar r^{2}\over \ell^{2}}\right)\,,
\eeq
which, for large $\bar r$, yields
\beq
\beta_{0}=\beta\left[1+{\ell^{2}\over 2\bar r^{3}}(L_{c}-L)\right]\,,
\eeq
where $L_{c}=0$ for $k=1,0$ and it is defined by eq.\ \eqref{mcrit} for $k=-1$. Thus, the Euclidean action for the black hole is
\beq
I_{E}^{bh}=\int d^{4}x\sqrt{g}R^{2}=\int_{0}^{\beta}d\tau\int_{r_{+}}^{\bar r}drr^{2}R^{2}={48\beta\over \ell^{4}}(\bar r^{3}-r_{+}^{3})\,.
\eeq
By combining the expressions above, the terms containing $\bar r$ cancel out and  we finally obtain
\beq
\Delta I_{E}={24\beta\over \ell^{4}}\left( -r_{+}^{3}+k\ell^{2}r_{+} \right)-{48\beta L_{c}\over \ell^{2}}\,,
\eeq
with the usual understanding that $L_{c}=0$ for $k=1,0$. Now, with the help of eqs.\ \eqref{thermo}, we finally find the energy and the entropy of the black hole, expressed respectively by
\beq\label{energy}
E={48\over \ell^{2}}(L-L_{c})\,,
\eeq
and
\beq\label{entropy}
S={96\pi r_{+}^{2}\over \ell^{2}}.
\eeq
Note that the expression of the energy is in line with the results found in \cite{deser}. Note also that  this formula agrees with Wald's prediction eq.\ \eqref{wald}, provided one replace $X(R_{H})$ with $|X(R_{H})|$ so that the sign is positive\footnote{Recall that in our case $X=2R\sim -\ell^{-2}$.}. This is reasonable since the origin of this term can be traced back to the conformal transformation between Jordan and Einstein frames. The conformal factor reads precisely $X(R)$ but it must be definite positive in order to preserve the metric signature. From this, we see that the absolute value is necessary and, as a byproduct, Wald's formula is correct also when $X(R)$ is negative-definite.
 
It is instructive to compare the formulae above with the ones found in GR, which read   \cite{topbh}
\beq
E_{GR}=M-M_{c}\,,\quad S_{GR}={A\over 4G}\,,
\eeq
where $G$ is the Newton's constant, $A$ is the horizon area, and $M$ is the physical mass of the hole. We note immediately that, in contrast to the GR case, both energy and entropy  depend also on the anti-de Sitter radius $\ell$. In addition, the entropy does not depend on $G$, as expected since the scale invariance of the action for quadratic gravity does not allow  dimensionful parameters.

Our final goal is to write down the first law of black hole thermodynamics. In GR we know that this has the universal form $TdS_{GR}=dE_{GR}$ for all $k$. In our case we find instead that
\beq
TdS=dE+{48 (L-L_{c})\over \ell^{3}}d\ell=dE+{E\over\ell}d\ell\,,\quad \forall\,\,\, k\,,
\eeq
where $E$ and $S$ are the quantities defined in eqs.\ \eqref{energy} and \eqref{entropy}, while $T=\beta^{-1}$. As expected, the first law takes in account the variations of the parameter $\ell$ since it is as much as arbitrary as $L$. 
Is the last contribution a pressure term? We may answer noting that we can rewrite the above first law as
\beq
TdS=dE+\frac{E}{3  V}\,d V\,,
\eeq
where we have set $V= \ell^3$. In this form, this equation clearly shows that the black hole thermodynamics is described by the same equation governing  a gas of massless radiation, namely
\beq
P=\frac{E}{3 V}\,.
\eeq
This result is fully consistent with the scale invariance of the model. In fact, by using the the expression for $E$, we can also write, for all $k$
\beq
\ell \,TdS=48\,d(L \,\ell^{-1})\,,
\eeq
which shows that the only relevant parameter of the theory is the dimensionless  ratio $L/ \ell$, since also the temperature is scale invariant. This becomes manifest if we write
\beq
\ell\, T={3z^{2}+k\over 4\pi z}\,,
\eeq
where we introduced the dimensionless parameter 
\beq
z={r_{+}\over \ell}\,,
\eeq
related to $L/\ell$ by
\beq
{L\over \ell}=kz+z^{3}\,.
\eeq

 We now compute the  heat capacities, formally defined as
\beq
{\cal C}_{V}=T\left(\partial S\over \partial T\right)_{V}\,,\quad {\cal C}_{P}=T\left(\partial S\over \partial T\right)_{P}\,.
\eeq
In terms of $z$ we find, for all $k$
\beq
{\cal C}_{V}={192\pi z^{2}(3z^{2}+k)\over (3z^{2}-k)}\,,
\eeq
and
\beq
{\cal C}_{P}={768\pi z^{2}(3z^{2}+k)(z^{3}+kz-z_{0})\over z(z^{2}-k)(3z^{2}+5k)-4z_{0}(3z^{2}-k)}\,,
\eeq
where $z_{0}=k(1-k)/(3\sqrt{3})$. As expected, these quantities are all scale invariant. By inspection, we find that ${\cal C}_{P}(k=1)>0$ for $z>1$, while ${\cal C}_{P}(k=-1)>0$ when $z>\sqrt{3}/ 3$, which corresponds to $r_{+}>r_{c}$.

Finally one can study the $PV$ and $PT$ phase diagrams by using the explicit expression of $P$ in function of $V$ and $T$
\beq
P(V,T)={16\over 27V^{4/3}}\left[(9+4\Delta^{2}+9k)\Delta+\sqrt{\Delta^{2}+3k}(9+4\Delta^{2}+3k) \right]\,,
\eeq
where
\beq
\Delta=\sqrt{4\pi^{2}T^{2}V^{2/3}-3k}\,.
\eeq
Concerning the isobaric $PT$ curves, we find that $P$ is a monotonically growing function of $T$ for all $k$. In the case of isothermal $PV$ curves, $P$ is a monotonically decreasing function of $V$ only for $k=0$ or $k=-1$. This is the behavior found in many standard homogeneous thermodynamical systems.
For $k=1$ instead, the pressure has a global maximum for $V_{max}=(\pi T)^{-3}$. This is consistent with the fact that, for any fixed pressure $P<P(V_{max})=2(2\pi T)^{4}$, there are two black holes with different size. 
 
\subsection{The thermodynamical state space}

The above considerations bring to attention the structure of the thermodynamical state space, and the corresponding identification of the meaningful physical quantities of the theory.  

Let us first  consider the parameter $k$ as fixed. The thermodynamical degrees of freedom are thus encoded in any triplet of quantities like, for example, $(r_{+},L,\ell)$ or $(P,V,T)$, related by an equation of state. To be definite, let us discuss the theory in terms of the parameters $(L,\ell)$, which form a two-dimensional vector space. Physical states are not single points, since two metrics related by a scale transformation 
\beq
ds^2\to\lambda^2ds^2\,,
\eeq
are both solutions of the field equations and physically indistinguishable in a scale invariant theory. Thus, we adopt the interpretation of scale  invariance as the property that all scaled metrics  describe the same physics (or that they are physically equivalent, remember that only ratios of space-time intervals are meaningful).   Under an arbitrary scaling, we obtain a family of metrics differing only by the flow
\beq
(L,\ell)\to(\lambda L,\lambda \ell)\,.
\eeq
Every point on this line (i.e.\  an orbit of the dilatation group) defines a class of equivalent metrics. So, inequivalent physical states will be in one-to-one correspondence with straight lines in $(L,\ell)$ space, in agreement with the  well-known fact that a continuous mass spectrum is not at odds with scale invariance.   

If we accept only positive scaling (i.e. $\lambda>0$), on the ground that we should not admit negative mass black holes, then the state space (namely, the space of orbits) is clearly homeomorphic to a circle, and each solution is characterized by a given value of the positive ratio $L/\ell$. The quantities that are constant on the orbits are the physically meaningful ones; such is the entropy, the ratio $L/\ell$ and the heat capacities $C_{P}$ and $C_{V}$. 

The restriction to positive values of $\lambda$ has a striking consequence:  each orbit comes arbitrarily close to the origin at $L=\ell=0$. Looking at the metrics we see that they approach a kind of $AdS_{2}$ metric: more precisely, in the limit $(L,\ell)\rightarrow (0,0)$ (with $L/\ell=$ const ), and at any finite $r$,  all metrics tend to the line element
\beq
ds^2=-{r^2\over\ell^2}dt^2+{\ell^2\over r^2}dr^2+r^2d\omega^2\,.
\eeq
By rescaling the coordinates according to $r\rightarrow \ell /Y$ and $t\rightarrow \ell /T$, one finds
\beq
ds^2=\left(\ell^{2}\over Y^{2}\right)\Big[-dT^2+dY^2+d\Sigma^2_{k}\,\Big]\,,
\eeq
which is manifestly conformal to Minkowski space for $Y\neq 0$. Alternatively, the above metric can be seen as the direct product  $AdS_{2}\times{\cal H}_{2}^{k}$, namely the two-dimensional anti-de Sitter space times the horizon space manifold. This is the only solution which belongs to all orbits simultaneously, so it is tempting to consider it as a universal ground state for all solutions discussed so far (this is the second possibility that we mentioned in subsection \ref{ssA}).

It is interesting to see what happens if we include also scalings with negative $\lambda$. The space of orbits is then the projective two-dimensional space $RP^2$ and all orbits intersect again at the point $AdS_{2}\times{\cal H}_{2}^{k}$ of the previous case. This leads to the bizarre consequence that we should now consider states with both $L<0$ and $\ell<0$ to be physically equivalent to the more familiar ones with positive $L$ and $\ell$. However,  these solutions  are not all black holes: depending on $k$, some are wormholes and some other have naked singularities. We do not know whether this is a permitted extension of the formalism, nor whether it makes any physical sense. For example, the toroidal black hole would be dual (in the sense of being on the same orbit) to a naked singularity, with no horizon, no temperature but some positive entropy. The physical interpretation of these solutions is not clear but it certainly deserves further investigation.

So far, we have considered $k$ as a fixed and discrete parameter.  In classical gravity it is an accepted wisdom that no topology changing-process exists. Thus, within this framework, each $k$ operates like a superselection parameter and the state space discussed above retains its form. On the other hand, there could be topology-changing amplitudes whenever the path integral between initial and final  configurations with different $k$ is non-vanishing. In this case, $k$ is not a superselection parameter and the phase space structure of the theory is more complicated. For example, if there were a solution of the Euclidean equations of motion which interpolates between a toroidal and a spherical topology, there could be such a topology-changing process. We leave this possibility as an open question.

\subsection{Asymptotically flat black holes}\label{flatbh}

For this class we encounter major differences with the corresponding cousins in GR. Since the Ricci scalar vanishes, both the volume part as well as the boundary terms vanish when evaluated on the solutions, leading to a vanishing partition function. At the same time, the Wald entropy vanishes because $X(R)=2 R=0$, and the total energy vanishes too by the BHW theorem \cite{Boulware:1983td}. So the first law takes the  trivial form $``0=0''$, and there are formally no contradictions. On the other hand the surface gravity on the horizon is certainly finite, suggesting the existence of a well-defined horizon temperature. Black holes with no entropy but finite temperature are not new, see e.g.\ \cite{cai23}.  In the present context, we have two suggestions. 

One is that in pure $R^2$ gravity there is nothing to radiate but thermal gravitons; however, gravitons can be consistently defined only in a perturbative sense, and it may well be the case that the equations governing the perturbations are not scale invariant, since they require a choice of background that breaks scale invariance. There is a well-known example in condensed matter physics: the Euler equations of the fluid are Galilei invariant although the equations for the sound waves are not (they are  formally Lorentz invariant!) \cite{unruh}.  More generally, adding matter makes the black hole radiate. But in doing so, we break again the scale invariance, and ordinary thermodynamics is promptly recovered. 

The other suggestion is that there is graviton radiation and the theory is still scale invariant but the quantum state  is pure. Here too we have a famous example: the Schwinger pair production process has an effective temperature although the state of the emitted radiation is pure. Both possibilities seem sound, therefore we cannot draw any definitive conclusions regarding the physical interpretation of these black holes. 

\section{Concluding remarks}

\noindent We have given an almost exhaustive description of black hole thermodynamical states in $R^2$ gravity, taking into account the special role played by the scale symmetry of the action. Solutions are partitioned into classes with fixed values of some ratios; in particular, no fixed mass or temperature can be assigned to a black hole without specifying also the anti-de Sitter radius. This has the consequence that, in formulating thermodynamical laws, this radius must also be varied along with the mass parameter, in such a way as to give scale invariant laws. In fact the pressure could have been predicted in this way. We have found that the entropy is always a meaningful quantity, which is constant within each class and coincides, for all cases, with the Noether charge formula proposed by Wald, up to a necessary change of sign. In fact, from the thermodynamical point of view, each class of solutions can be labeled uniquely by its entropy. Moreover, in the absence of topology changing processes, the discrete parameter $k$ acts as a superselection charge, making the different topologies to behave as separate worlds. If topology changing processes exist, this is evidently no longer true.

The above analysis may be extended to the generic quadratic scale invariant gravity model
\beq
{\cal L}=\sqrt{g}\left( c_1 R^2+ c_2 R_{\mu \nu} R^{\mu \nu}+ c_3   R_{\mu \nu \alpha \beta}R^{\mu \nu \alpha \beta} \right) \,.
\eeq
It is well-known that only two out of the three terms of the action are really independent, since one of them can be eliminated by making use of the Gauss-Bonnet quadratic invariant, which does not contribute to the classical equation of motion.  
A direct computation shows that  one has a class of static, topological spherically symmetric solution with the asymptotically (anti) de Sitter metric \eqref{gtt2}. Thus, the analysis of the nature  of pure $R^2$ gravity can be repeated here without any modification. 
Note, however, that in the computation of the entropy with the Wald method, the Gauss-Bonnet terms give a non vanishing contribution. 

We also observe that with the specific choice  $c_1=\frac{1}{3}$, $c_2=-2$, and $c_3=1$, one is dealing with the so-called Weyl conformal gravity, where the Lagrangian density is $C_{\mu \nu \alpha \beta}C^{\mu \nu \alpha \beta}$, namely a  quadratic invariant of the Weyl tensor   $C_{\mu \nu \alpha \beta}$. This model has been extensively studied, and static black hole solutions have been found \cite{r,m,m2}. The corresponding topological black hole solutions have been investigated in 
\cite{klem,guido,pope,guido12}. 

Finally, we recall that a longtime ago Buchdahl realized that every Einstein space with arbitrary cosmological constant is a solution of the equations of motion of $R^{2}$ gravity \cite{buchdahl}. It follows that also the Kerr-Newman-(anti) de Sitter metrics with arbitrary mass, angular momentum, charge, and (anti) de Sitter radius is a solution to $R^{2}$ gravity. The thermodynamical properties of these black hole solutions can be investigated along the lines drawn here  and they will be the subject of a future paper.

\end{document}